# A Survey On Various Data Mining Techniques for ECG Meta Analysis

**Kratika Tyagi**
Department of CSE,
ASET, Amity University
NOIDA, Uttar Pradesh, India
Email ID: kratika.tyagi@student.amity.edu

**Sanjeev Thakur**
Department of CSE,
ASET, Amity University
NOIDA, Uttar Pradesh, India
Email ID: sthakur3@amity.edu

*Abstract*— **Data Mining is the process of examining the information from different point of view and compressing it for the relevant data. This data can also be utilized to build the incomes. Data Mining is also known as Data or Knowledge Discovery. The basic purpose of data mining is to search patterns which have minimal user inputs and efforts. Data Mining plays a very crucial role in the various fields. There are various data mining procedures which can be connected in different fields of innovation. By using data mining techniques, it is observed that less time is taken for the prediction of any disease with more accuracy. In this paper we would review various data mining techniques which are categorized under classification, regression and clustering and apply these algorithms over an ECG dataset. The purpose of this work is to determine the most suitable data mining technique and use it to improve the accuracy of analyzing ECG data for better decision making.**

## I. INTRODUCTION

Data Mining is the process of examining the information from different point of view and compressing it for the relevant data. This data can also be utilized to build the incomes. Data Mining is also known as Data or Knowledge Discovery. The basic purpose of data mining is to search patterns which have minimal user inputs and efforts. Data Mining plays a very crucial role in the various fields. The objective of data mining procedure is to concentrate data from a dataset and change it into a justifiable structure for further utilization. There are various data mining techniques which can be applied in various fields of technology. By using data mining techniques, it is observed that less time is taken for the prediction of any disease with more accuracy.

Electrocardiography (ECG) is the procedure of recording the electrical activity of the heart over a timeframe using electrodes set on a patient's body. These electrodes recognize the minor electrical changes on the skin that rise up out of the heart muscle depolarizing in the midst of each heartbeat.

Biomedical applications have played an imperative role in the enhancement of medical diagnosis and have provided many solutions for the identification of diseases. The primary aim of such applications is to help doctors and medical practitioners make effective decisions by analyzing various computer generated reports. However in most cases it has been observed that there is a huge difference in what is interpreted and what is enacted. Using data mining techniques we aim to minimize the error of making effective decisions by categorizing various Interpretations of ECG data. This paper will have a proper literature survey of the various data mining techniques over the ECG data to find the most suitable classifiers and clustering algorithms. This will be followed by developing a data mining model and compare it with the most suitable data mining algorithm over ECG data.

## II. TECHNIQUES USED IN DATA MINING

**A.** *Association*

Association is considered as the best data mining strategy. In this mining strategy, pattern is considered on the basis of a connection of a specific thing or assortment of things in the same exchange. Association method is generally utilized in the prediction of the heart diseases as it let us know the relationship among different set of attributes used in analysis.

B. *Clustering*

Clustering is another data mining techniques which makes clusters of objects which are similar in characteristics. Clustering basically defines classes and put objects in them. Example- By using clustering technique to predict the heart disease we get clusters or we can make the list of those patients which are having the same risk factor and separate list for those patients which are having high blood pressure.

C. *Classification*

Classification is a data mining strategy which depends on machine learning. Classification is basically utilized to classify things in a specific arrangement of information into one of predefined set of classes or groups. It uses numerical methods like decision trees, neural networks etc.

D. *Prediction*

Prediction is another data mining technique which defines the relationship between reliant and autonomus variables. Example- Prediction test can be utilized as a part of anticipating the benefit for future as for deal. If we consider benefit as a reliant variable and offer as autonomus variable. At that point we can draw a relapse bend which could be utilized for predicting the benefit based on the historical data of offer and benefit.

III. RELATED WORK

Here in this section, all papers are discussed which are taken into account for the review process.

In [1], Keeley Crockett et al presented an approach which combines the multiple decision trees and utilizes the power of fuzzy inference techniques and a back proportion feed forward neural network which improves the overall classification. It has been concluded that back proportion feed forward neural network has a better classification accuracy as compared to multiple crisp decision trees.

In [2], Anju Rathee et al explained various algorithms such as Iterative Dichotomiser 3 (ID3), C4.5 and Classification and regression Tree (CART) and compared their performance and results and their evaluation is done by the existing datasets. The performance of different decision tree algorithms are investigated on the basis of their certainty and time seized for inferring the tree. Among all the three algorithms discussed in this paper, it has been observed that C4.5 is the best algorithm among all since it gives the preferred precision and proficiency over the other algorithms.

In [3], Xianen Qiu et al discussed about Sparse Fuzzy C-Means method in view of inadequate grouping system. Sparse Fuzzy C-Means (SFCM) algorithm is proposed by the author which usually reforms traditional Fuzzy C-Means to deal with higher data dimension cluster which depends on Witten's inadequate clustering structure. Feature Selection is embedded from Sparse Fuzzy C-Means into Fuzzy C-Means via inadequate weighting and thereby generate the understanding of the model much easier. After having the comparison, it has been observed that this technique can choose the vital elements and thereby increases the performance for higher cluster problems.

In [4], Amany Abdelhalim et al proposed a strategy named as RBDT 1 (Rule Based Decision Trees) which utilizes an arrangement of interpretive rules as the suggestion for generating the decision tree. RBDT 1 is compared with the ID3 method and it has been concluded that RBDT 1 performs superior than the other strategies with respect to the complexities.

In [5], Zarita Zainuddin et al described that Fuzzy C-Means generally partitions the observations into various clusters which are normally based on the principles of fuzzy theory. Moreover Euclidean distance minimization in Fuzzy C-Means tends to detect the hyper-spherical shaped clusters which is not feasible for the real world problems. So an effective Fuzzy C-Means algorithm is proposed which adopts the symmetry similarity measure in order to search the appropriate clusters. It has been observed that several artificial and real life data sets applied on different nature and performance assessment with other clustering algorithms demonstrate its superiority.

In [6], Navjot Kaur et al focussed on K-means clustering and its algorithm. The unproved outcome of K-means Clustering and execution of K-means with respect to execution time is discussed. Beside this, K-Means clustering algorithm have a limitation too, that it requires more space for execution. In order to minimize the run time, Ranking Method is used. It is further demonstrated that how clustering is achieved in fewer execution time related to other common strategies. It has been concluded that it makes an endeavor of examining the attainability of K-Means clustering strategy utilizing Ranking Method as a part of data mining.

In [7], Amir Ahmad et al presented a method called random projection random discretization ensembles which fundamentally makes the gatherings of multivariate decision trees with the help of univariate decision tree algorithm. Random Discretization is proposed which basically creates random discretized features. It has been concluded that RPRDE is used basically in small ensembles and it is also robust to the noisy data.

In [8], G.V. Nadiammai et al discussed about intrusion detection as it has become the useful tool against the network attack as it allows network administrator to detect vulnerability. In this paper, the performance of the three clustering algorithms are explained namely, K-means, Fuzzy C-means, Hierarchichal algorithms over intrusion data set and time complexity are recorded. It has been concluded that Fuzzy C-Means algorithm executes superior comparatively with rest two algorithms with respect to performance and time.

In [9], A.M. Chandrashekar et al proposed a contemporary methodology for network intrusion detection system which is a combination of different data mining strategies such as Fuzzy C means, Neural Network and Support vector machine. This methodology decreases the quantity of characteristics connected with information focuses utilizing the Neural Network. For execution assessment measurements comprising of five parameters such as affectability, specificity, exactness, precision and F value are used. The assessment is built on the basis of training and testing for all intrusions namely DOS, PROBE, R2L and U2R and it is concluded that proposed technique works better and achieve the best results among the rest technique.

In [10], Jyoti Yadav et al discussed standard k-means algorithm and analyzed its shortcoming. It also discuss three dissimilar algorithms which removes its limitations and improves the speed and efficiency which results in optimal number of clusters. The first algorithm evacuates the restriction of determining the estimation of k. The second algorithm reduces the computational complexity furthermore evacuates the issue of dead unit. The third algorithm diminishes the time complexity. It has been concluded that the time complexity of first algorithm is greater than standard k-means algorithm for the larger data set. So if the data structure is used to store the information in first algorithm then the time complexity of that algorithm can be reduced which will result in optimal solution.

In [11], P.Viswanath et al focused on the enhancement to the k-NNC known as k-nearest neighbor mean classifier (kNNMC) which discovers k-nearest neighbors for every class of the training pattern and furthermore discover means for all k-neighbors. It has been concluded that proposed strategy can be executed in coordinate frame and incorporated with any space reduction and has been shown experimentally with the help of standard datasets which demonstrates better arrangement over k nearest neighbor classifier, weighted k nearest neighbor classifier and k-NNC using Hamamoto's bootstrapped training set.

In [12], Thomas A. Runkler et al focussed on partially supervised clustering. Two approaches namely fuzzification and harmonic means are examined for making the k means model lesser sensitive. According to the BIRCH data set, it has been concluded that k-harmonic means is less delicate to initialization when contrasted with k means and delivers better clustering results.

In [13], Niphat Claypo et al proposed an assessment on the surveys of Thai cafeteria with the assistance of K-means Clustering and MRF Feature Selection which reduces the number of features. It has been inferred that K-means clustering is appropriate with MRF Feature Selection as it achieves the best execution.

IV. REVIEW OF VARIOUS ALGORITHMS STUDIED

*(A). DECISION TREES*

A Decision Tree is a structure which joins the source nodes, subsidiary and leaf nodes. Every interior nodes demonstrates an evaluation on a property, every subsidiary represents the evaluation of a test, and each leaf node holds the class name. The highest node in the tree is the source node.

*Advantages of decision trees*

1. Decision trees are distinct and when they are compacted they are simpler to follow.
2. Decision trees can deal with both nominal and numerical data traits.
3. Decision tree illustrartion is easier to produce the estimation of any discrete classifier.
4. Decision trees can deal with data sets which may have errors.

*Disadvantages of decision trees*

1. When divide and conquer method is used, decision trees perform well if only few attributes are available. But if many complex interactions are present then it does not work well.
2. Decision trees are not stable as a little bit change in information can bring about extensive change in the tree.
3. Decision trees are inclined to mistakes in classification.
4. Preparing decision trees with numerous branches are complicated and tedious.

*(B). K MEANS*

K means clustering is a data mining/machine learning strategy utilized to organize information into gatherings of related perceptions without information of those connections. The k means strategy is one of the easiest clustering strategy which is generally utilized in medical imaging, biometrics and associated field.

*K means advantages*

1. When huge variables are present then k means works faster than hierarchical clustering.
2. When the clusters are globular then k means produce tighter cluster as compared to hierarchical clustering.

*K means disadvantages*

1. It is difficult to predict the k value.
2. It does not work well with global cluster.
3. It also not works well with clusters which are having different densities and different size.

*(C). K NEAREST NEIGHBORS*

K nearest neighbor algorithm is non parametric strategy which is utilized in both the classification and regression. K-nn is a type of occurrence based learning where the task is approximated locally and all the calculation is conceded until grouping.

*Advantages of K nearest neighbors*

1. The expenses for the learning process is zero.
2. No assumptions about the characteristics of the ideas to learn must be finished.
3. Complex ideas can be learned by local approximation using easy method.

*Disadvantages of K nearest neighbors*

1. It is costly to find k nearest neighbors when the dataset is very large.
2. Performance depends on the number of dimensions.

*(D). NAIVE BAYES*

The Naive Bayes algorithm relies on the contingent probabilities. It utilizes Bayes theorem an equation that registers the likelihood by checking the recurrence of qualities and groupings of qualities in the recorded information.

*Advantages of Naive Bayes*

1. The Naive Bayes strategy is quick and versatile model.
2. The build process for Naive Bayes is parallelized.
3. Naive Bayes can be utilized as a part of paired and multiclass grouping issues.

*(E). FUZZY C-MEANS*

Fuzzy C-Means is an information clustering strategy in which a dataset is assembled into n groups with every information point in the dataset fitting with each group to a particular degree.

*Advantages of Fuzzy C-Means*

1. It gives best result for the overlapped dataset and it is better than k means algorithm.
2. In fuzzy c-means information is allotted membership to every cluster center where data point can has a place with more than one group focus.

*Disadvantages of Fuzzy C-Means*

1. Apriori specification of the number of clusters.
2. Euclidean distance measures can unequally weight hidden variables.

## V. CONCLUSION

The basic purpose of data mining is to search patterns which have minimal user inputs and efforts. In this paper we have discussed various data mining techniques which are categorized under classification, clustering, prediction, association and based on those techniques we reviewed different research papers and we will apply these algorithms over an ECG data set. The purpose of this work is to determine the most suitable data mining technique and we found that classification is better technique comparing with other, so we will use it further to improve the accuracy of analyzing ECG data for better decision making.